# NMR lineshape of $^{29}$Si in single-crystal silicon

Brooks Christensen and John C. Price*
*Department of Physics, University of Colorado, Boulder, CO 80309, USA*
(Dated: September 25, 2016)

We report measurements of the NMR lineshape of 4.685% abundant $^{29}$Si in single-crystal silicon for four different crystallographic orientations relative to the applied magnetic field. To avoid significant inhomogeneous broadening, the sample crystals are immersed in a susceptibility matched liquid and the proton NMR lineshape in the liquid is used to measure the residual susceptibility mismatch and to shim the applied field. The observed lineshapes are in good agreement with disorder-averaged spin dynamics simulations performed on a $4 \times 4 \times 4$ unit cell lattice with periodic boundary conditions. The splitting of resolved doublet features is used to measure the asymmetry $\Delta J$ of the nearest-neighbor indirect dipole or J-coupling tensor, with the result $\Delta J = 90 \pm 20$ Hz. All resolved doublets can be identified with specific nearest and next-nearest neighbor isolated spin pairs that occur in the disorder ensemble.

## I. INTRODUCTION

Coherent many-body quantum dynamics of nuclear spins in solids is an old subject, dating back to observations of the free-induction-decay in CaF$_2$, a cubic array of spin-1/2 $^{19}$F nuclei coupled by magnetic dipole interactions.[1,2] Coherent spin transport has been observed directly in the same system,[3] and multi-pulse echo techniques have enabled observations of the spread of many-spin correlations in CaF$_2$ and other extended systems dominated by magnetic-dipole interactions.[4–7]

New interest in nuclear spin dynamics in solids, particularly in single-crystal silicon and diamond, arises from proposals and experiments in quantum information processing.[8–10] In some proposals, the qubits are spin-1/2 $^{29}$Si nuclei in crystalline silicon and dipole coupling is used to implement two-qubit operations.[11,12] Achieving adequate coherence may depend on removing unwanted spins by isotopic purification,[8] and/or by active decoupling methods.[13] Accurate models of spin-spin interactions and many-body dynamics are essential to these ideas. Spin transport by dilute spins in solids is also relevant to hyperpolarization for quantum information processing and medical imaging applications.[14,15]

First-principles theory for many-body quantum dynamics in extended spin systems is very challenging, and the dependence of observables on geometry, dimensionality, and disorder is poorly understood. The NMR free-induction-decay is the most tractable of the observable phenomena. When a strong Zeeman field is present, the initial thermal equilibrium state involves only single-spin operators. After a 90° pulse, multi-spin correlations begin to appear in the density matrix, but the magnetization may decay completely before a large number of spins become correlated. If the number of correlated spins needed to describe the signal is not greater than 10–20, the problem may be simulated on a classical computer. For the ordered cubic geometry of CaF$_2$, a short-time expansion[16] gives an excellent account of the free-induction-decay, as the number of terms required in the expansion corresponds to the number of correlated spins that must be included.

In this work, we observe and analyze the NMR lineshape (Fourier transform of the free-induction-decay) of single-crystal silicon, an extended dipole-coupled nuclear spin system with well-defined disorder. Natural silicon consists of two stable spin-0 isotopes, $^{28}$Si and $^{30}$Si, and one stable spin-1/2 isotope, $^{29}$Si. The natural abundance of $^{29}$Si is 4.685%, so that single-crystal silicon is an ordered lattice with that fraction of the sites randomly occupied by NMR-active $^{29}$Si nuclei. The known natural abundance is the only parameter needed to specify the disorder in the system. The two Si sites in the face-centered-cubic silicon lattice have the same chemical shift (Zeeman energy) because they are related by an inversion symmetry of the lattice. Thus, the rotating-frame Hamiltonian of the system contains only magnetic dipole interactions between pairs of identical spin-1/2 nuclei. These interactions vary in strength according to bond length and orientation.

The NMR lineshape in single crystals is often broadened by inhomogeneity of the applied magnetic field. Field distortions are not so serious in 100% abundant $^1$H and $^{19}$F systems, which usually have dipolar linewidths of order 10 kHz. However, all candidates for isotopically disordered systems (i.e. those elements with a mixture of stable spin-1/2 and spin-0 isotopes) have much lower gyromagnetic ratios, will typically be at lower concentration, and will thus have much smaller intrinsic linewidths. Linewidths reported here are roughly 80 Hz, or 1 ppm at our 99.31 MHz Larmor frequency. This is comparable to diamagnetic susceptibilities of pure liquids and solids. Thus, if no special care is taken, susceptibility mismatch will produce inhomogeneous broadening comparable to the intrinsic linewidth, as was the case for previous lineshape measurements of $^{29}$Si in single-crystal silicon.[17–19] Moreover, in solid-state NMR there may be no narrow reference lines suitable for shimming (adjusting field homogeneity). We describe a method below in which the sample crystal is immersed in a susceptibility-matched liquid. The proton lineshape in the liquid is used both to shim the field and to determine the residual field inhomogeneity within the crystal.

Our lineshape results are compared to spin simulations on a 128-site lattice in which spins are randomly placed at natural abundance. The simulated spectrum is averaged over 100 disorder configurations for each crystal orientation. The predicted spectrum is in very good agreement with our measurements, but only if a nearest-neighbor indirect dipole interaction (J-coupling tensor) term is included in the Hamiltonian together with the direct dipole interaction. Doublet features in the spectrum are identified with specific spin configurations in the disorder ensemble.

## II. EXPERIMENT

### A. Samples

The samples were made from (110) oriented float-zone silicon wafers,[30] 50 mm diameter, 0.525 mm thick and with specified resistivity of 5 kΩ cm to 10 kΩ cm. The wafers were diced into 2.6 mm × 5.0 mm rectangular chips, with the longer edge of the chip oriented relative to flats that were provided on the wafers. On four different wafers, the longer edge was oriented in each of the four directions shown in Fig. 1. These became the magnetic field directions when the chips were mounted in the spectrometer. Orientation of the chips was checked by X-ray diffraction using the rotating crystal method to ensure that no gross errors were made, but final alignment was relative to the wafer flats, which were specified accurate to ±0.5°.

For each orientation, a rectangular prism sample 2.6 mm × 2.6 mm × 5.0 mm was assembled from a stack of five chips using very small drops of epoxy. Each sample was attached, again with a small drop of epoxy, to an 0.33 mm outside-diameter capillary aligned with the longer dimension. The whole assembly was slipped into a standard 5.0 mm diameter NMR tube. The capillary served to keep the sample at the height of the probe coil and to keep the longer dimension of the sample aligned with the tube axis. With all sources of error included, we believe that the four crystallographic directions shown in Fig. 1 were aligned with the magnetic field to ±2°.

### B. Susceptibility Matching

All NMR measurements were done in a 500 MHz Varian Unity spectrometer using 5 mm diameter silicate glass NMR tubes. A Varian broadband probe with an inner X-nucleus coil and an outer proton/deuterium coil allowed for observations of protons near 500 MHz and $^{29}$Si near 100 MHz.

To avoid field distortion, the single-crystal silicon samples were immersed in a susceptibility-matched liq-uid. The diamagnetic susceptibility of silicon, $\chi_{Si} = -3.4 \times 10^{-6}$, is relatively small in magnitude. (We use dimensionless SI susceptibilities in this paper.) Because many solvents are more diamagnetic than silicon, they can be matched by adding a paramagnetic solute. To prepare a starting liquid, we first filled a 5 ml volumetric flask with 80% fully-deuterated acetone (by volume) and 20% protonated acetone. We then added 51.5 mg paramagnetic chromium(III) acetylacetonate [Cr(acac)$_3$] and two drops of tetramethylsilane (TMS) shift reference to the flask. This produced a liquid that was less diamagnetic than silicon, i.e. the Cr(acac)$_3$ concentration was too high. An NMR tube with a silicon sample as described above was filled so that the liquid extended 2.5 cm above and below the center of the silicon crystal. The tube was then inserted into the spectrometer, and the proton singlet from the acetone was observed and shimmed.

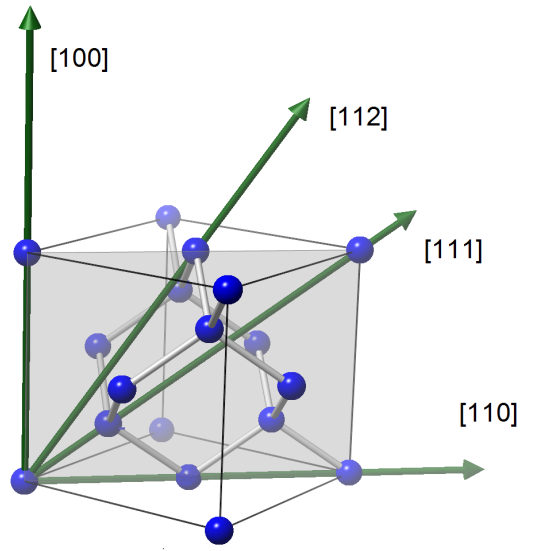

FIG. 1: Silicon conventional cubic unit cell. The samples were oriented with the applied magnetic field along the four directions shown. Each direction is in a (110) plane.

A typical example of the lineshape of the starting liquid is shown in Fig. 2(a) (black curve). The main peak is similar to what was observed in a tube filled with only the liquid. The linewidth is about 5 Hz, due to relaxation from the Cr(acac)$_3$. There is also a second smaller peak 50 Hz above the main peak, which moves to lower frequency if the Cr(acac)$_3$ concentration is lowered by adding acetone to the tube. By adding small amounts of acetone to the tube, and repeatedly observing the lineshape, the smaller peak can be made to merge with the main peak, resulting in the lineshape shown in Fig. 2(b) (black curve).

To understand this behavior, we simulated the lineshape due to field distortion using the boundary-element magnetics package RADIA.[20,21] The liquid volume within the probe coil was modeled as a cylinder 4.0 mm in diameter and 20 mm long, with the 2.6 mm × 2.6 mm × 5.0 mm crystal centered on the cylinder and aligned with the cylinder axis. Both the liquid and solid regions were magnetized along the cylinder axis in proportion to their susceptibilities. The simulated magnetic field was sampled at 224,000 points within the simulation volume. The component parallel to the axis was converted to frequency units, histogrammed and line-broadened with a 5 Hz Lorentzian to include the effects of relaxation. The simulated proton lineshape is shown in Fig. 2(a) (blue curve) for a susceptibility difference $\Delta\chi = \chi_{Si} - \chi_{liquid} = -0.92 \times 10^{-6}$ and in Fig. 2(b) (blue curve) for a susceptibility difference $\Delta\chi = \chi_{Si} - \chi_{liquid} = -0.10 \times 10^{-6}$. By inspecting the simulated magnetic field, we find that the smaller peak is due to return flux in the liquid volume between the crystal and the inner wall of the tube. With $\Delta\chi$ negative, the crystal is more diamagnetic than the liquid, so the induced field within the crystal is anti-parallel to the applied field and the return flux is parallel to the applied field. Thus the smaller peak appears at a higher frequency than the main peak. The simulated main peak in Fig. 2(a) also shows a tail to lower frequency, due to the opposing induced field above and below the crystal. The tail is less prominent in the observed lineshape than in the simulation, probably because the shim fields are able to partly compensate this distortion. However, the smaller peak is a robust feature whose position provides an in-situ measurement of the susceptibility difference between the crystal and the liquid. When the small peak merges with the main peak, the liquid is susceptibility matched to the crystal.

The red curves in Fig. 2 show the simulated lineshapes for $^{29}$Si within the crystal, due to contributions from field distortion only. The RMS widths for the cases illustrated are 8.6 Hz for the starting fluid and 0.93 Hz for the matched fluid, assuming $\Delta\chi$ has the values given in the figure caption. Based on proton lineshapes measured for each sample, corresponding simulations, and similar simulations aimed at estimating field distortion due to small glue drops, we conclude that the inhomogeneous broadening of $^{29}$Si within the crystal is in all cases less than 3 Hz RMS. This susceptibility matching approach can be used for other diamagnetic solids as long as their susceptibility is not below about $-14 \times 10^{-6}$, beyond which no sufficiently diamagnetic solvents are available.

### C. $^{29}$Si Lineshape

Because of the very long longitudinal relaxation time $T_1$ of undoped silicon, data collection methods were chosen to avoid unnecessary recovery intervals. The $^{29}$Si signal from TMS in the liquid was used to find the 90° pulse width, which was always close to 15 μs. $T_1$ in the crystal was determined by first accurately zeroing the polarization with a sequence of three 90° pulses, then waiting a repolarization time from 0.5 h to 23 h before recording the free induction decay after a 90° pulse. This sequence was repeated for various repolarization times and the integrated signal intensity versus repolarization time was fit to an exponential to find $T_1$. This method is insensitive to small errors in the 90° pulse width and a good estimate of $T_1$ can be obtained from only a few data points. $T_1$ values in the range 5 h to 7 h were found for all orientations.

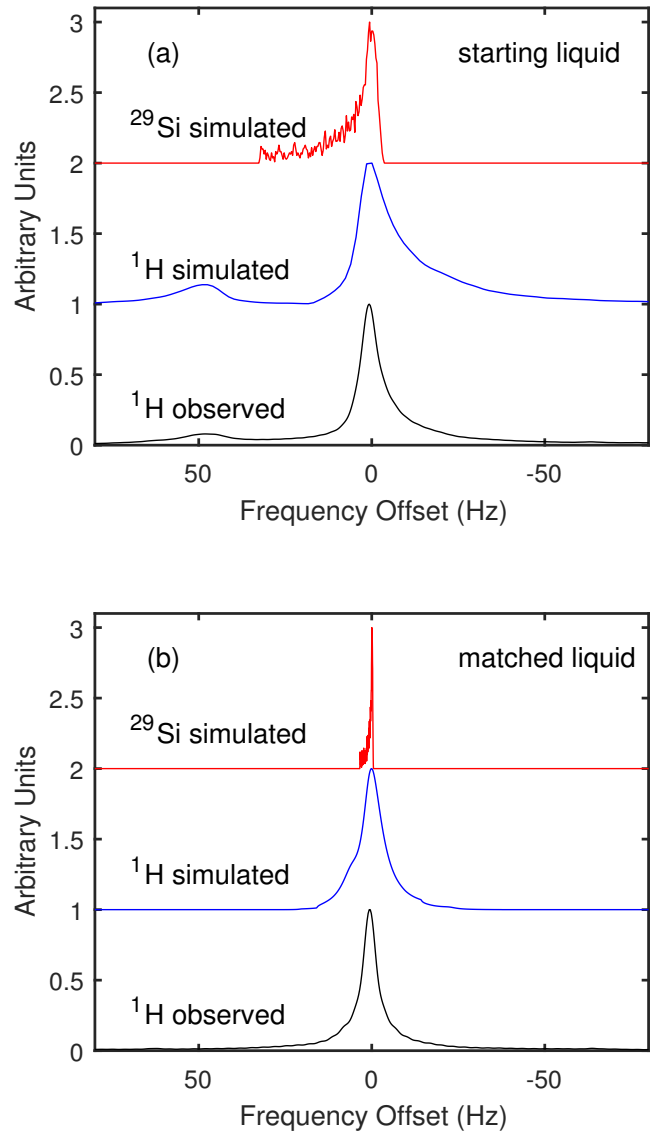

FIG. 2: Observed (black) and simulated (blue) $^1$H lineshape in susceptibility-matching liquid; simulated (red) $^{29}$Si lineshape in solid due to field inhomogeneity. (a) Starting liquid, lineshapes simulated with susceptibility difference $\Delta\chi = \chi_{Si} - \chi_{liquid} = -0.92 \times 10^{-6}$. (b) Matched liquid, lineshapes simulated with $\Delta\chi = -0.10 \times 10^{-6}$.

A calculation was done to find the best data collection strategy to maximize the signal-to-noise ratio in a given data collection period. For $T_1 = 5$ h and a 24 h data collection period, the optimal choice is to use a 70° pulse and a 4 h recovery time. Each free-induction-decay (FID) signal was recorded at a 50 kHz sampling rate for one second. The total amount of data collected was 216, 380, 240, and 164 h for the [100], [112], [111] and [110] orientations respectively.

For the results presented here, the FIDs are multiplied by $\exp(-t/\tau)$ with $\tau = 32$ ms (10 Hz exponential apodization), Fourier transformed, individually phased, centered, and baseline corrected before all of the spectra for each orientation are summed. In some cases where not all individual spectra have equal integrated intensity, the sum is weighted by the integral to maximize the final signal-to-noise ratio. The resulting lineshapes are shown in Fig. 3 (black curves). The spectra are plotted relative to their center frequencies, which are close to 99.314 MHz. The chemical shift of $^{29}$Si in silicon is found to be -82.5 ppm relative to the TMS $^{29}$Si signal in the matching liquid. There is a 2.1 kHz wide background signal at -112 ppm due to silicon in the sample tube and the probe structure. This background is sufficiently displaced from the silicon crystal signal so that it does not interfere with our measurements.

## III. ANALYSIS

### A. Hamiltonian

The general NMR Hamiltonian for a homonuclear spin-1/2 system (in angular frequency units) is[22,23]

$$\begin{aligned} H = & -\sum_i \gamma S_i^z B^z \\ & -\sum_{i<j} b_{ij} \left( 3 \frac{(\vec{S}_i \cdot \vec{r}_{ij})(\vec{S}_j \cdot \vec{r}_{ij})}{r_{ij}^2} - \vec{S}_i \cdot \vec{S}_j \right) \\ & +2\pi \sum_{i<j} \vec{S}_i \mathbf{J}_{ij} \vec{S}_j \;, \end{aligned} \quad (1)$$

where the three terms are the Zeeman interaction with an applied magnetic field in the $\hat{z}$ direction, the direct magnetic dipole interaction, and the tensor J-coupling or indirect magnetic dipole interaction. The dipole coupling constant is given by

$$b_{ij} = \frac{\mu_0 \gamma^2 \hbar}{4\pi r_{ij}^3} \;. \quad (2)$$

The gyromagnetic ratio for $^{29}$Si is

$$\gamma = -53.190 \times 10^6 \,\mathrm{rad\,s^{-1}\,T^{-1}} \;, \quad (3)$$

and $r_{ij}$ is the length of the coordinate vector $\vec{r}_i - \vec{r}_j$ extending from spin $j$ to spin $i$. We omit the small chemical shift part of the Zeeman interaction which is isotropic and equal for every spin in the silicon lattice.

Because the $^{29}$Si Larmor frequency is much greater than the ($\approx 1$ kHz) nearest-neighbor dipole interaction, we may make the usual secular approximation. This is done by first transforming to the interaction representation with respect to the Zeeman term, and then dropping all terms in the transformed Hamiltonian with coefficients that oscillate at the Larmor frequency or at twice the Larmor frequency.

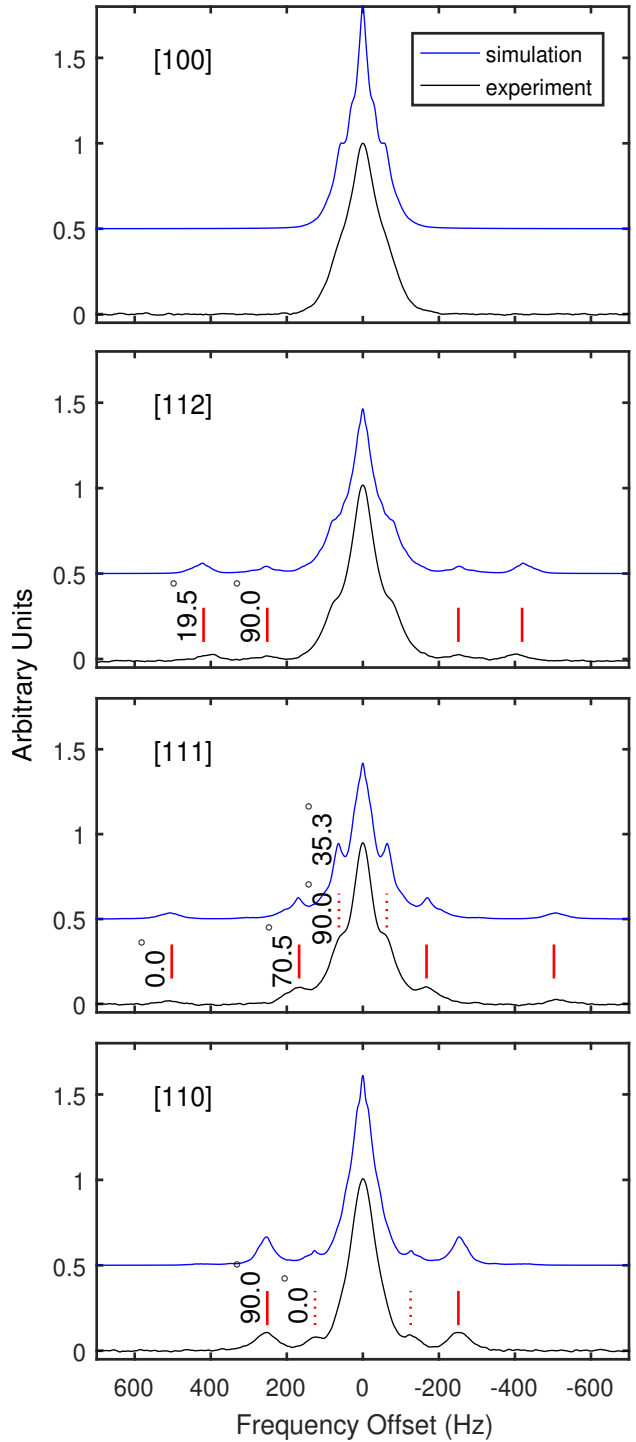

FIG. 3: Observed (black) and simulated (blue) lineshape for the four crystal orientations. The J-coupling parameters are $J = 70$ Hz and $\Delta J = 90$ Hz. Red lines indicate doublet splittings for isolated spin pairs, evaluated with Eqn. 7 and discussed in Sec. III C. Solid red lines are for nearest-neighbor pairs and dotted red lines are for next-to-nearest-neighbor pairs.

The terms involving $\mathbf{J}_{ij}$ can be simplified if we take advantage of symmetries. As far as we are aware, no results have been reported in the literature for the tensor elements $\mathbf{J}_{ij}$ in silicon crystals. In general, the J-coupling is expected to be weaker than the direct dipole interaction and to fall off rapidly with bond order. Therefore, we consider here only the nearest-neighbor J-coupling. In the silicon structure, the lattice has an inversion center at the midpoint of each nearest-neighbor bond, the bond is a 3-fold rotation axis, and 3 mirror planes intersect along the bond. Because of these symmetries, the nearest-neighbor $\mathbf{J}_{ij}$ is a symmetric matrix[24] and it has axial symmetry with one principle axis along the bond and two equal axes perpendicular to the bond. Switching temporarily to Cartesian coordinates with $\hat{z}$ along the bond, we have

$$\mathbf{J} = \begin{bmatrix} J_{xx} & 0 & 0 \\ 0 & J_{yy} & 0 \\ 0 & 0 & J_{zz} \end{bmatrix} , \quad (4)$$

and $J_{xx} = J_{yy}$. It is conventional to describe an axial J-coupling tensor in terms of $J$ and the asymmetry $\Delta J$, defined as

$$\begin{aligned} J &= \frac{1}{3}(J_{xx} + J_{yy} + J_{zz}) \\ \Delta J &= J_{zz} - \frac{1}{2}(J_{xx} + J_{yy}) . \end{aligned} \quad (5)$$

With this simplification, the secular Hamiltonian becomes

$$\begin{aligned} H =& -\sum_{i<j} b_{ij} \left(\frac{3\cos^2\theta_{ij}-1}{2}\right) \left(3 S_i^z S_j^z - \vec{S}_i \cdot \vec{S}_j\right) \\ &+ 2\pi {\sum_{i<j}}' \frac{\Delta J}{3} \left(\frac{3\cos^2\theta_{ij}-1}{2}\right) \left(3 S_i^z S_j^z - \vec{S}_i \cdot \vec{S}_j\right) \\ &+ 2\pi {\sum_{i<j}}' J \vec{S}_i \cdot \vec{S}_j , \end{aligned} \quad (6)$$

where the primes on the second and third sums indicate that they are to be taken over nearest-neighbor pairs only, and $\theta_{ij}$ is the angle between the magnetic field direction $\hat{z}$ and $\vec{r}_i - \vec{r}_j$. We find that the asymmetry $\Delta J$ has an effect equivalent to changing the strength of the direct dipole interaction between nearest neighbors.

In Sec. III C below we relate features in the spectrum to specific pairs of nearby spins. For this task, it is helpful to have an explicit formula for the splitting $\Delta f_{ij}$ (in Hz) of a doublet due to an isolated pair of spins. When only two spins are present, $\vec{S}_i \cdot \vec{S}_j$ commutes with $S_i^z S_j^z$ and with the initial density operator representing uniform polarization. As a consequence, the splitting depends only on the $S_i^z S_j^z$ terms in Eqn. 6 and is given by

$$\Delta f_{ij} = \left| \left( \Delta J - \frac{3}{2\pi} b_{ij} \right) \left( \frac{3\cos^2\theta_{ij}-1}{2} \right) \right| , \quad (7)$$

for nearest neighbors. For isolated pairs other than nearest neighbors, the same formula applies with $\Delta J = 0$.

### B. Spin Dynamics Simulations

We attempt to describe our data by numerically computing the FID of finite clusters of spins representing members of a disorder ensemble. To generate a member of the ensemble, spins are placed at random on a silicon lattice containing $4 \times 4 \times 4$ primitive unit cells and 128 silicon sites. The number of spins in a member of the ensemble is Poisson distributed and has a mean of 6.0 at the natural abundance of 4.685%. For the main results presented here, a disorder ensemble with 100 members containing a maximum of 11 spins is used. The probability of an ensemble member containing more than 11 spins is 2%, so if the ensemble is regenerated it will sometimes contain members with more spins. Computation time increases very rapidly with the number of spins, so such ensembles are not used. This choice slightly biases our results towards low density.

For the initial state, we use the thermal density operator in the high temperature limit

$$\rho(0) = \frac{1}{2^N}\left(\mathbf{I} + \frac{\hbar \omega_L}{kT} M^z\right) , \quad (8)$$

where $N$ is the number of spins, $\omega_L$ is the Larmor frequency, and $M^z$ is the z-component of the total spin operator

$$\vec{M} = \sum_i \vec{S}_i . \quad (9)$$

After a 90° pulse, the initial deviation density operator $M^z$ is converted to $M^x$, and then it propagates under the Hamiltonian Eqn. 6. The FID is computed as the expectation value of $M^x$:

$$\langle M^x(t) \rangle = \mathrm{Tr}\left(M^x \exp(-iHt) M^x \exp(iHt)\right) . \quad (10)$$

Simulations are implemented using the spin dynamics package SPINACH,[25,26] which includes features that improve the speed and accuracy of our calculations. We use the spherical tensor operator basis set and restrict the basis[27] to include only coherence order +1 and -1. This restriction does not introduce any approximation because only these coherence orders are present in the initial state and the secular Hamiltonian Eqn. 6 commutes with $M^z$ and therefore conserves coherence order. SPINACH includes features for further restricting the basis set in various ways, but the main results reported here include all product operators (with coherence order +1 and -1) and are essentially exact. The Hamiltonian and density matrix are stored in sparse form. The Liouville-space propagation algorithm in SPINACH uses the Krylov method to improve efficiency and includes features that ensure numerical accuracy. For the dipole interactions, a 15 Å cut-off distance is used and contributions due to periodic boundary conditions are evaluated by periodically extending the lattice twice in each direction along each lattice vector. Periodic boundary conditions are also used for the nearest-neighbor J-coupling interactions.

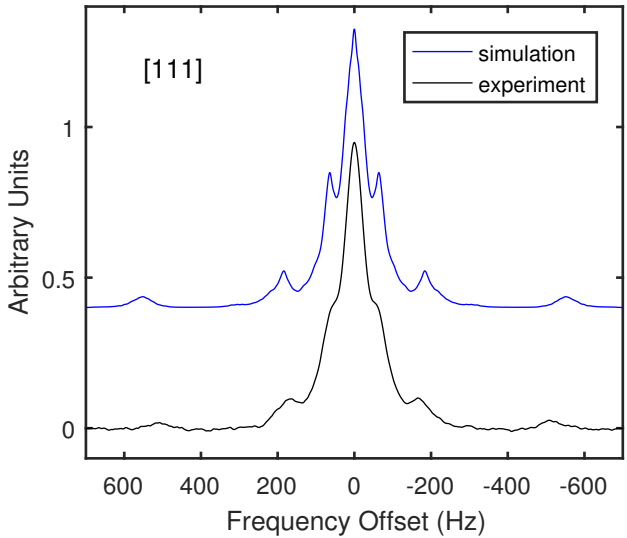

FIG. 4: Observed (black) and simulated (blue) lineshape for [111] orientation, with J-couping parameters $J = \Delta J = 0$.

For each member of the disorder ensemble, the spin system is propagated and the FID is stored at 0.2 ms intervals for a total time of 0.1 s. Before the individual FIDs are averaged together, they are normalized so that the initial amplitude is proportional to the number of spins in the ensemble member. After averaging, the FID is apodized with a 10 Hz exponential ($\tau = 32$ ms), and then Fourier transformed. This is the same processing that is applied to generate spectra from the experimental FIDs. All experimental and simulated spectra are normalized to have the same integral over frequency, corresponding to the fact that they all represent the same spin density.

Figure 4 shows the experimental (black) and simulated (blue) results for [111] orientation of the crystal, with the J-coupling parameters $J = \Delta J = 0$. All qualitative features of the lineshape are reproduced in the simulation, but it is evident that the wide doublet with a splitting of 1005 Hz in the experiment is too wide in the simulation. In Sec. III C below, we show that this doublet is due to a specific nearest-neighbor pair in the disorder ensemble. According to Eqn. 7 above, the splitting of an isolated nearest-neighbor pair includes a term proportional to $\Delta J$. Repeating the simulation with $\Delta J = 90$ Hz brings it in line with experiment, as shown in Fig. 3. We estimate that $\Delta J$ is determined with an accuracy of $\pm 20$ Hz. Note from Eqn. 7 that positive $\Delta J$ will reduc-ing the splitting magnitude for any value of $\theta_{ij}$, as long as the J-coupling contribution is small compared to the direct dipole contribution.

Liquid-phase measurements[28,29] suggest that the isotropic J-coupling parameter is likely to be in the range 50 Hz to 100 Hz. The simulations reported here are insensitive to the value of $J$ within this range. This is not surprising, because $J$ does not influence the splitting of an isolated nearest-neighbor pair, and it can only be significant for a three-spin configuration where two are nearest neighbors and the third is in close proximity. Such configurations are rare in the disorder ensemble. For the main results reported here we use $J = 70$ Hz.

The experimental lineshapes (black curves) and corresponding simulations (blue curves) for all four crystal orientations are compared in Fig. 3 above. The only significant adjustable parameter in the simulations is $\Delta J$. The widths and shapes of the central peaks agree well with the simulations, as do the locations and widths of the resolved doublet peaks. We emphasize that the simulations involve unitary evolution without relaxation parameters; the resulting disorder-averaged FIDs have only been line-broadened by 10 Hz in the same way as the experimental data. The results are not sensitive to the exact line-broadening used because all real features in the data are more broad than 10 Hz. There is a tendency in the simulations for some doublet peaks and also the feature at the very center of the line to be somewhat too sharp. In addition, the simulated doublet splittings in [112] orientation are slightly too wide. These discrepancies are discussed further below.

In Fig. 5 we show [111] orientation lineshape simulations for three statistically independent disorder ensembles with 30, 100, and 300 members. The 100 member case is the same simulation as in Fig. 3 above. In the 300 member case the inner doublet peaks are less prominent than in the 100 member case. This suggests that our simulations are not quite converged with respect to the size of the disorder ensembles.

Simulated lineshapes for lattices containing $3 \times 2 \times 2$, $3 \times 3 \times 3$, and $4 \times 4 \times 4$ primitive unit cells are shown in Fig. 6. The simulations are for [111] orientation and in each case the disorder ensemble contains 100 members. The number of sites for these cases are 24, 54, and 128 and the corresponding mean number of spins in the ensemble members are 1.13, 2.53, and 6.00. We see a very rapid change in the lineshape as lattice size is increased, and a marked reduction in the sharpness of the peaks, especially at the center of the line. It is likely that our simulations are not fully converged with respect to lattice size.

The features that are too sharp in the simulations in comparison with experimental data may be due in part to the simulated lattice size being too small and in part to the ensembles being too small. The simulations presented here took about 100 hours on a 6 core, 3.5 GHz desktop processor. It seems likely that larger scale spin simulations could significantly improve the agreement.

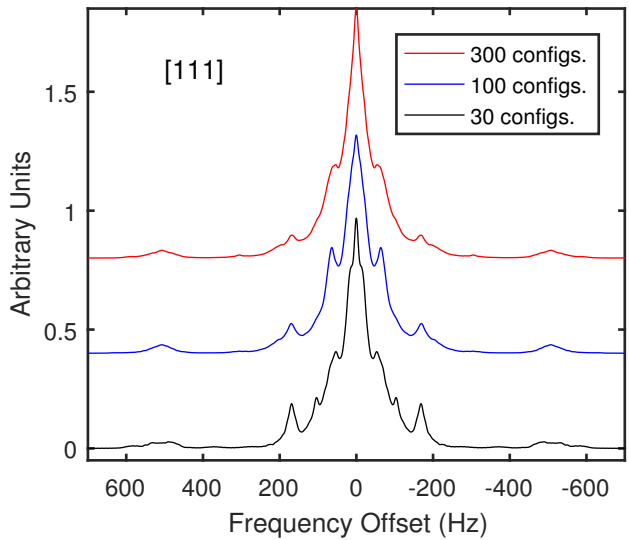

FIG. 5: Simulated lineshape in [111] orientation for disorder ensembles with 30, 100 and 300 members. The J-coupling parameters are $J = 70\,\text{Hz}$ and $\Delta J = 90\,\text{Hz}$.

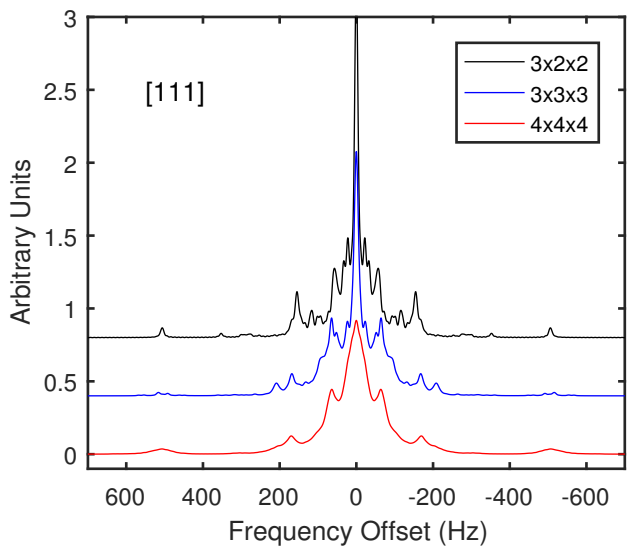

FIG. 6: Simulated lineshape in [111] orientation for three different lattice sizes. Each case is simulated with a disorder ensemble with 100 members and with J-coupling parameters $J = 70\,\text{Hz}$ and $\Delta J = 90\,\text{Hz}$.

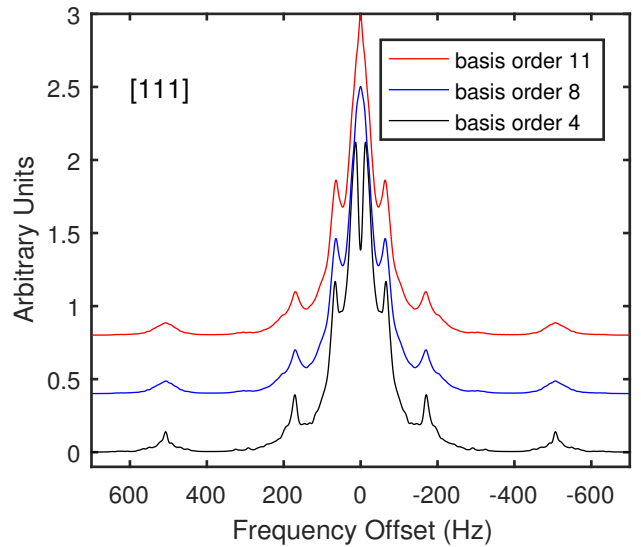

FIG. 7: Simulated lineshape in [111] orientation for three different basis sets. The basis sets contain all states with products of up to $P$ spin operators, where $P$ is 4, 8 or 11. Each case is simulated with a disorder ensemble with 100 members and with J-coupling parameters $J = 70\,\text{Hz}$ and $\Delta J = 90\,\text{Hz}$.

One may ask if the basis set we are using might be larger than it has to be. In Fig. 7 we show simulations in [111] orientation with the basis set restricted to only include products of up to $P$ single-spin operators, where $P$ is 4, 8, or 11. The case $P = 11$ is exact because no member of the ensemble contains more than 11 spins. For $P = 8$, members of the ensemble with 8 or fewer spins are simulated exactly, but members with more spins are simulated with a restricted basis containing all products of up to 8 spin operators. When the basis is restricted, the Liouvillian is also modified so that the system will remain in the restricted subspace under time evolution, ensuring that evolution remains unitary.[25] We see that products of up to 4 operators is not a large enough basis, but restriction to products of up to 8 operators is almost as good as an exact calculation.

### C. Structure Identification

Following Ref. 17, we can associate the wide resolved doublets in Fig. 3 with specific isolated pairs that occur in the disorder ensemble.

In [111] orientation, 1 nearest-neighbor bond is at $0°$ relative to the applied field, producing the largest possible splitting. As discussed above, the J-coupling asymmetry $\Delta J = 90\,\text{Hz}$ was chosen so that the 1005 Hz splitting for this isolated pair, obtained from Eqn. 7, would agree with the data. The other 3 nearest neighbors are at $70.5°$ with a predicted splitting of 335 Hz, which agrees well with the data. The calculated splittings are indicated by the red lines in Fig. 3. They also agree well with our many-spin simulation, showing that the positions of these resolved peaks can be understood by considering only isolated pairs. In [111] orientation, 6 next-to-nearest neighbors are at $90.0°$ and 6 are at $35.3°$. All 12 have the same predicted splitting of 126 Hz, as indicated by the dotted red lines in the figure. Although not fully resolved, these pairs correspond well with the prominent shoulders on the central peak.

In [112] orientation, there is 1 nearest neighbor at $19.5°$, 1 at $90.0°$, and 2 at $61.9°$, giving predicted splittings of 838 Hz, 503 Hz, and 167 Hz respectively. The wider two doublets are clearly visible in the data and in the simulation, but the predicted splittings for the wider doublet appears about 40 Hz too wide. This is likely due to alignment error. If the bond angle is increased from 19.5° to 21.5°, the splitting from Eqn. 7 for the wider doublet agrees with the data. An error of this size is consistent with our ±2° alignment uncertainty. Alignment uncertainty does not contribute significantly to the determination of $\Delta J = 90$ Hz from the widest [111] doublet, because in that case the bond angle is 0° and Eqn. 7 is insensitive to small orientation errors. The pairs in [112] orientation at 61.9° may contribute to the shoulder on the central peak, but there are several other next and next-to-next nearest neighbor bond orientations with about the same splitting.

For [110] orientation, all 4 nearest neighbor bonds are at 90.0°, with predicted splitting of 503 Hz, in good agreement with the data and simulation. The narrower doublet appears to correspond to 2 of the 12 next-to-nearest neighbors at 0° with predicted splitting of 252 Hz.

Finally, in [100] orientation, all nearest-neighbor bonds are at the magic angle 54.7° and no wide doublets appear. Four next-nearest neighbors at 90.0° and 4 next-to-next nearest neighbors at 25.2° may contribute to the shoulder on the central peak.

## IV. SUMMARY AND CONCLUSIONS

We present measurements of the NMR lineshape of 4.685% abundant $^{29}$Si in single-crystal silicon for 4 different crystal orientations. To avoid any significant inhomogeneous broadening due to distortion of the applied field by the sample susceptibility, we immerse the sample in a susceptibility matched liquid. The proton NMR lineshape in the liquid is used to directly measure the susceptibility difference between the liquid and the sample, to shim the applied field, and to infer an upper bound of 3 Hz on inhomogeneous broadening for $^{29}$Si in the samples.

The observed lineshapes are compared to spin dynamics simulations averaged over a disorder ensemble, which is populated by placing spins at random in a $4 \times 4 \times 4$ unit cell lattice with periodic boundary conditions. The Hamiltonian includes direct dipole interactions and an axisymmetric nearest-neighbor J-coupling tensor. The simulations are insensitive to the isotropic $J$ parameter, but the splitting of wide doublets due to isolated nearest-neighbor spin pairs provides a measurement of the asymmetry parameter $\Delta J = 90 \pm 20$ Hz. The simulated lineshapes are in good agreement with the data, reproducing all major features with no adjustable parameters other than $\Delta J$, and doing so without any relaxation parameters.

Resolved doublet features can be associated with specific nearest-neighbor and next-nearest-neighbor isolated spin pairs that occur in the disorder ensemble. All such doublets are identified, as are some prominent shoulders on the central peaks. By choosing a doublet with a bond angle of 0° to measure $\Delta J$, the result becomes insensitive to small crystal alignment errors.

The methods described here might be applied to better understand the NMR lineshape of other isotopically disordered crystals containing $^{29}$Si, as well as those containing other dilute spin-1/2 nuclei such as $^{13}$C, $^{77}$Se, and $^{207}$Pb.

## V. ACKNOWLEDGMENTS

The authors thank Arthur Pardi and Richard Shoemaker for help with the spectrometer. We also thank Rahul Nandkishore, Ana Maria Rey, Charles Rogers, Ilya Kuprov and Jonathan Yates for helpful comments. This work was supported by an IGP seed grant from the University of Colorado, and by the Interdisciplinary Center for Electronics at the University of Colorado.